\documentclass[epsfig,12pt]{article}

\usepackage{epsfig}

\newcommand{\be}{\begin{equation}}
\newcommand{\ee}{\end{equation}}
\newcommand{\ba}{\begin{eqnarray}}
\newcommand{\ea}{\end{eqnarray}}
\newcommand{\bd}{\begin{displaymath}}
\newcommand{\ed}{\end{displaymath}}

\def\thalf{{\textstyle{\frac{1}{2}}}}

\title{{\bf Equation of State and Phase Fluctuations near the
Chiral Critical Point}}
\author{{J. I. Kapusta} \vspace*{0.1in}\\
{\it School of Physics and Astronomy, University of Minnesota}\\
 {\it Minneapolis, Minnesota 55455, USA}}

\date{}

\begin{document}

\maketitle

\begin{abstract}
The thermodynamics and critical exponents and amplitudes of high temperature and dense matter near the chiral critical point are studied. The parameterized equation of state matches on to that calculated with lattice QCD at zero chemical potential and to the known properties of nuclear matter at zero temperature. The extent to which finite size effects wash out the phase separation near the critical point is determined. 
\end{abstract}

\vspace{0.5cm} {PACS: 25.75.Nq; 05.70.Jk; 21.65.Mn}

\newpage

\section{Introduction}

The up and down quark masses are very small but not zero.  In consequence, the conventional wisdom is that there is no true thermodynamic chiral phase transition at finite temperature $T$ and zero baryon chemical potential $\mu$.  Instead, there is expected to be a curve of first-order phase transition in the $\mu$-$T$ plane that terminates in a second-order phase transition at some critical point $(\mu_c,T_c)$.  The location of the critical point is obviously of quite some interest.  This topic has been under intense theoretical study using various effective field theory models, such as the Namu Jona-Lasinio model \cite{asakawa89}-\cite{scavenius01}, 
a composite operator model \cite{barducci}, a random matrix model \cite{halasz98}, a linear $\sigma$ model \cite{scavenius01}, an effective potential model \cite{hatta02}, and a hadronic bootstrap model \cite{antoniou02}, as well as various implementations of lattice QCD \cite{fodor02}-\cite{gavai05}.  
Reviews of this subject were presented by Stephanov \cite{stephanov} and Mohanty \cite{MohantyQM}.  

This subject is also of great interest because collisions between heavy nuclei at medium to high energy, such as at the future Facility for Antiproton and Ion Research (FAIR), or possible low energy runs at the Relativistic Heavy Ion Collider (RHIC), may provide experimental information on the phase diagram in the vicinity of a critical point.  One characteristic signature would be large fluctuations in phase space of conserved quantities, such as charge, baryon or strangeness, on an event by event basis \cite{Shuryak,Hatta}.  The variance of the distributions is proportional to the spatial size of the correlated region, which could be rather small due to the finite size and lifetime in heavy ion collisions \cite{small}.  This led to the suggestion to measure higher moments to search for non-Gaussian behavior \cite{nongaussian}.  In order to study these effects quantitatively, not only are measurements needed but also dynamical simulations of phase separation and fluctuations in heavy ion collisions \cite{Randrup}. 

The goal of this paper is to understand the basic features of the equation of state near the QCD chiral critical point and the magnitude of phase fluctuations in its vicinity.  The essential requirements are to incorporate the critical exponents and amplitudes, and to match on to lattice QCD results at $\mu = 0$ and to nuclear matter at $T = 0$.  We will accomplish this by parameterizing the Helmholtz free energy as a function of temperature and baryon density so as to incorporate the above requirements.

This work is similar to the study of the nuclear liquid gas phase transition in \cite{Goodman}.  An obvious difference is that the latter studied a transition between nuclear liquid and gas, whereas the present paper will study the transition between quark matter and hadronic matter.  Apart from that, this paper will develop a description of the equation of state that has the correct critical exponents which are not integers or simple fractions; in other words, it is not a mean field theory.  It will also include the critical amplitudes, which are universal.  The chiral phase transition is in the same universality class as the 3D Ising model and liquid gas phase transition.  Finally, the parameterization will incorporate knowledge about the zero baryon density equation of state, computed by lattice QCD, as well as knowledge about the high density behavior of nuclear matter at zero temperature.  Perhaps the closest work that addressed these issues was ref. \cite{attract} which blended a parameterization of the 3D Ising model equation of state into quark and hadron equations of state.  In that work there was an ambiguity as to how to relate the magnetization to the baryon chemical potential.  In this paper there is no such issue.  These two parameterizations can perhaps be viewed as alternatives which provide some idea as to the range of uncertainty in how to describe matter near the chiral critical point.

The outline of the paper is as follows.  In Sec. 2 we summarize some thermodynamic relations and definitions of the relevant critical exponents.  In Sec. 3 we develop a relatively general parameterization of the equation of state in the vicinity of the critical point whose motivation comes from mean field theory. In Sec. 4 we fix the parameters to best match onto what is known about the equation of state at finite temperature but zero baryon density, and at finite baryon density but zero temperature.  In Sec. 5 we show the numerical results obtained with the developed parameterization and parameters.  In Sec. 6 we apply the Landau theory of fluctuations away from the stable thermodynamic phases to estimate the magnitude of density fluctuations, which turn out to be surprisingly large.  Concluding comments are made in Sec. 7.  Those wishing to know only the results should read Sec. 5 and the Appendix which summarizes the parameterized equation of state.

\section{Thermodynamic Relations}

Consider the equation of state of matter in the vicinity of a critical point which is in the same universality class as the liquid-gas phase transition and the 3D Ising model.  It is advantageous to work with the Helmholtz free energy density $f$ which is a function of the temperature $T$ and baryon density $n$.  Some useful thermodynamic relations involving the pressure $P$, baryon chemical potential $\mu$, entropy density $s$ and energy density $\epsilon$ follow.
\ba
f &=& f(n,T) \\
P &=& n^2 \frac{\partial}{\partial n} \left( \frac{f(n,T)}{n}\right)\\
\mu &=& \frac{\partial f(n,T)}{ \partial n}\\
s &=& - \frac{\partial f(n,T)}{ \partial T}\\
\epsilon &=& f(n,T) + Ts(n,T)
\ea
Clearly these satisfy the thermodynamic identity $\epsilon = -P +Ts + \mu n$.  The heat capacity per unit volume $c_V$ and isothermal compressibility are
\ba
c_V &=& T \frac{\partial s(n,T)}{\partial T}\\
\kappa_T^{-1} &=& n \frac{\partial P(n,T)}{\partial n} \, .
\ea
The baryon number susceptibility is $\chi_B = n^2 \kappa_T$.  In what follows we shall focus on $\kappa_T$ rather than $\chi_B$ since they are so directly related to each other.

When discussing a critical point with critical temperature $T_c$ and baryon density $n_c$ it is useful to define
\ba
t &=& \frac{T - T_c}{T_c}\\
\eta &=& \frac{n - n_c}{n_c} \, .
\ea
The critical exponents $\alpha, \beta, \gamma, \delta$ are defined as follows.   When $t \gg |\eta |$ and $t > 0$
\ba
c_V &\sim& t^{-\alpha} \\
\kappa_T &\sim& t^{-\gamma} \, .
\ea
Along the coexistence curve
\be
n_l - n_g \sim (-t)^\beta \, .
\ee
Along the critical isotherm
\be
P - P_c \sim |\eta|^{\delta} {\rm sign}(\eta) \, .
\ee
Mean field theories normally give $\alpha = 0$, $\beta = 1/2$, $\gamma = 1$, and $\delta = 3$, as we shall see.  Typical fluids are measured to have $\alpha \ll 1$, $\beta \approx 1/3$, $1.2 < \gamma < 1.3$, and $4 < \delta < 5$.  For experimental measurements and results see \cite{fluidexpts} and references therein.  The 3D Ising model has $\alpha = 0.11$, $\beta = 0.325$, $\gamma = 1.24$, and $\delta = 4.815$.  A good general reference is \cite{Zinnbook}.  We will now proceed to parameterize the equation of state near the chiral critical point of QCD at increasing levels of sophistication.

\section{Parameterizing the Equation of State}

As mentioned earlier, there are certain properties near a critical point that are identical for all theories within the same universality class.  Away from the critical point the equation of state depends on the details of the degrees of freedom and the interactions among them.  To proceed we will first briefly review  generic descriptions that arise from most if not all mean fields theories.  The results will be used to motivate a more sophisticated parameterization relevant for QCD.

\subsection{Mean field theories}  

By mean field theories it is meant that although interactions are included, correlations among the particles are not.  This usually results in thermodynamic variables scaling as rational powers of $\eta$ and $t$.  This can be represented by expanding $f$ in a Taylor series in $\eta$ about $\eta = 0$.
\be
f = \sum_{k=0} f_k(t) \eta^k
\ee
The coefficient functions $f_k(t)$ themselves may be expanded in a Taylor series about $t=0$ and they all have energy dimension 4.  The resulting pressure is
\ba
P &=& \sum_{k=0} P_k(t) \eta^k \nonumber \\
P_k &=& (k+1)f_{k+1}(t) + (k-1)f_k(t) \, .  
\ea
At the critical point both $\partial P(n,T)/\partial n = 0$ and 
$\partial^2 P(n,T)/\partial n^2 = 0$.  This implies that $P_1(0)=0$ and $P_2(0)=0$, or equivalently, $f_2(0)=0$ and $f_3(0)=0$.  Similarly the other thermodynamic variables may be obtained, such as
\be
\mu = \frac{1}{n_c} \sum_{k=0} (k+1) f_{k+1}(t) \eta^k
\ee
and
\be
s = -\frac{1}{T_c} \sum_{k=0} f^{\prime}_k(t) \eta^k
\ee
where the prime denotes differentiation with respect to $t$.

The simplest model is to set $f_k(t)=0$ for all $k>4$.  A quartic polynomial for the free energy is typical in particle and condensed matter physics.  This means that $P$ is quartic and $\mu$ is cubic in $\eta$, which allows for the usual S-shaped curves.  Phase coexistence requires that the pressures and chemical potentials of the two phases be equal for $t<0$.
\ba
P(\eta_l,t) &=& P(\eta_g,t) \\
\mu(\eta_l,t) &=& \mu(\eta_g,t)
\ea
The subscripts $l$ and $g$ stand for the liquid high density phase and the gaseous low density phase, respectively.  This determines $\eta_l(t) > 0$ and $\eta_g(t) < 0$ along the coexistence curve.  For this model
\be
2f_2 \left( \eta_l - \eta_g \right) + 
(3f_3 + f_2) \left( \eta_l^2 - \eta_g^2 \right) +
(4f_4 + 2 f_3) \left( \eta_l^3 - \eta_g^3 \right) +
3f_4 \left( \eta_l^4 - \eta_g^4 \right) = 0
\ee 
and
\be
2f_2 \left( \eta_l - \eta_g \right) + 3f_3 \left(
\eta_l^2 - \eta_g^2 \right) + 4f_4 \left( \eta_l^3
- \eta_g^3 \right) = 0
\ee 
Since both $f_2$ and $f_3$ vanish at the critical point, $f_4$ should not in order that $\eta_l$ and $\eta_g$ go to zero as $t \rightarrow 0$.  These equations apparently do not have any simple solution.  Therefore assume that $f_3(t)=0$, which is certainly not the most general case but it does allow us to gain valuable insight.  Then the solution to these equations is
\be
\eta_l(t) = -\eta_g(t) = \sqrt{\frac{-f_2(t)}{2f_4(t)}}
\ee
So the function $f_2(t)$ should be negative for $t<0$ and positive for $t>0$.  The simplest choice, usually obtained in mean field approximations, is that $f_2(t) \sim t$.  This gives $\beta = 1/2$.  The resulting coexistence curve in the $T$-$n$ plane is symmetric about $n_c$.  Real fluids oftentimes have an asymmetric curve.

Along the coexistence curve the chemical potential is
\be
n_c \mu_x(T) = f_1(t)
\ee
If, for example, the coexistence curve $T$ versus $\mu$ is parameterized as an ellipse then the function $f_1(t)$ is determined.  The coexistence pressure is
\be
P_x(T) = P_0(t) + \frac{f_2^2(t)}{4f_4(t)} =
- f_0(t) + f_1(t) + \frac{f_2^2(t)}{4f_4(t)}
\ee
Along the critical isotherm
\be
P(n,T_c) - P(n_c,T_c) = 4 f_4(0) \eta^3 + 3 f_4(0) \eta^4
\ee
so that the critical exponent $\delta = 3$.  The thermal compressibility is
\be
\kappa_T^{-1} = \frac{n}{n_c} \left[ P_1(t) + 2 P_2(t) \eta
+ 3 P_3(t) \eta^2 + 4P_4(t) \eta^3 \right]
\ee
When $t \gg |\eta | \rightarrow 0$ and $t > 0$
\be
\kappa_T \rightarrow \frac{1}{2f_2(t)}
\ee
so that if $f_2(t)$ vanishes linearly then the critical exponent $\gamma = 1$.  In the same limit
\be
c_V = - \frac{1}{T_c} f^{\prime \prime}_0(t)
\ee
If $f_0(t)$ is a regular function then the critical exponent $\alpha = 0$.
  
The limit of meta-stability is the isothermal spinodal.  It is determined by the condition $\partial P(n,T)/\partial n = 0$.  In this model one finds, with $f_3(t)=0$, that the lower limit is $\eta_1 = \eta_g/\sqrt{3} < 0$ and the upper limit is $\eta_2 = \eta_l/\sqrt{3} > 0$.  For $\eta_g < \eta < \eta_1$ the system is in a meta-stable gas phase, and for $\eta_2 < \eta < \eta_l$ the system is in a meta-stable liquid phase.

\subsection{A realistic parameterization}

Now we construct a model that has the correct critical exponents.  The most important consideration is to obtain the correct value of $\delta$ which is an irrational number.  Motivated by the mean field theories, we parameterize the free energy as
\be
f = f_0(t) + f_1(t)\eta + f_2(t)\eta^2 + f_{\sigma}(t)|\eta|^{\sigma} \,. 
\ee
The pressure, chemical potential, and entropy density are
\be
P = -f_0+f_1 + 2f_2\eta + f_2\eta^2
 + \sigma f_{\sigma} |\eta|^{\sigma - 1} \, {\rm sign}(\eta)
 + (\sigma - 1) f_{\sigma} |\eta|^{\sigma}
\label{pressure}
\ee
\be
n_c \mu = f_1 + 2f_2 \eta + \sigma f_{\sigma}
|\eta|^{\sigma - 1} \, {\rm sign}(\eta)
\label{chempot}   
\ee
\be
T_c s = -f^{\prime}_0 - f^{\prime}_1\eta - f^{\prime}_2\eta^2 - f^{\prime}_{\sigma}|\eta|^{\sigma} \, .
\label{entropy}
\ee
Phase coexistence is determined by equal pressures and chemical potentials at the same temperature but different densities.
\ba
2 f_2 \left( \eta_l - \eta_g \right) + \sigma f_{\sigma}
\left( |\eta_l|^{\sigma - 1} + |\eta_g|^{\sigma - 1} \right)
&=& 0 \\
f_2 \left( \eta_l^2 - \eta_g^2 \right) + (\sigma - 1)  f_{\sigma}
\left( |\eta_l|^{\sigma} - |\eta_g|^{\sigma} \right)
&=& 0
\ea
The second of these has an obvious solution for $\eta_g = - \eta_l$.  When substituted in the first equation we find
\be
\eta_l(t) = \left[ \frac{-2f_2(t)}{\sigma f_{\sigma}(t)} 
\right]^{\textstyle \frac{1}{\sigma-2}} \, .
\ee
Since $f_2(0)=0$ the pressure along the critical isotherm is
\be
P(n,T_c)-P(n_c,T_c) = \sigma f_{\sigma}(0) |\eta|^{\sigma-1} \, {\rm sign}(\eta)
+ (\sigma-1) f_{\sigma}(0) |\eta|^{\sigma}
\ee
and so the critical exponent $\delta = \sigma-1$.

The limit of meta-stability is the isothermal spinodal.  It is determined by the condition $\partial P(n,T)/\partial n = 0$, as mentioned earlier.  Now one finds that the lower limit is $\eta_1 = \eta_g/\delta^{1/(\delta - 1)} < 0$ and the upper limit is $\eta_2 = \eta_l/\delta^{1/(\delta - 1)} > 0$.  For $\eta_g < \eta < \eta_1$ the system is in a meta-stable gas phase, and for $\eta_2 < \eta < \eta_l$ the system is in a meta-stable liquid phase.  In the range $\eta_1 < \eta < \eta_2$ the system is unstable against isothermal fluctuations.

The density difference goes to zero as
\be
\eta_l - \eta_g \sim (-t)^{\beta} \, .
\ee
In the 3D Ising model and in real liquid-gas transitions it turns out that the thermal compressibility $\kappa_T$ diverges as $\kappa_+ t^{-\gamma}$ when $t \rightarrow 0^+$.  Since
\be
\kappa_T \rightarrow \frac{1}{2 f_2(t)}
\ee
when $\eta \rightarrow 0$ first, it follows that $f_2(t) \sim t^{\gamma}$ for $t \rightarrow 0^+$.  Putting these together, assuming that $f_2(t)$ has the same critical exponent for both positive and negative $t$, yields
\be
\gamma = \beta (\delta - 1)
\ee
which is a well-known relationship.  To allow for the possibility of asymmetry about $t=0$ we write $f_2(t) = \pm b_{\pm} \, (\pm t)^{\gamma}$, where the sign is chosen according to whether $t$ is positive or negative, and the $b_{\pm}$ are both positive numbers..

The heat capacity at $\eta \rightarrow 0$
\be
c_V \rightarrow - \frac{1}{T_c} f^{\prime \prime}_0(t)
\ee
diverges as $t^{-\alpha}$ when $t \rightarrow 0^+$.  Therefore we should write
\be
f_0(t) = {\bar f}_0(t) - a_+ t^{2-\alpha} 
\ee
as $t \rightarrow 0^+$ where ${\bar f}_0(t)$ is a smooth function.  Hence the singular part of $c_V$ is $c_+ t^{-\alpha}$ with $T_c \, c_+ = (2-\alpha)
(1-\alpha) a_+$.  Another relationship among the critical exponents is
\be
\alpha + 2\beta + \gamma = 2\
\ee
Once again we allow for asymmetry about $t=0$ and write
\ba
f_0(t) = \left\{ \begin{array}{ll}
\bar{f}_0(t) -a_- (-t)^{2-\alpha} & \mbox{if $t<0$} \\
\bar{f}_0(t) -a _+ t^{2-\alpha} & \mbox{if $t>0$ \, .}
\end{array} \right.
\ea

Using the best values arising from the 3D Ising model \cite{Guida} and from data on real liquid-gas phase transitions one has $\beta = 0.325$ and $\gamma = 1.24$.  The above relationships then imply $\delta = 4.815$ and $\alpha = 0.11$.  Note that the mean field model considered previously respects both of these relationships among critical exponents too.

Along the coexistence curve the chemical potential is
\be
n_c \mu_x(T) = f_1(t)
\ee
and the pressure is
\ba
P_x(T) &=& P_0(t) + \frac{\sigma - 2}{2} f_{\sigma}(t)
\left[\frac{-2f_2(t)}{\sigma f_{\sigma}(t)}\right]^{\frac{\sigma}{\sigma-2}}
\nonumber \\
&=& -f_0(t)+f_1(t) + \frac{\sigma - 2}{2} f_{\sigma}(t)
\left[\frac{-2f_2(t)}{\sigma f_{\sigma}(t)}\right]^{\frac{\sigma}{\sigma-2}} \, .
\ea   
The formula for the isothermal compressibility along the coexistence curve is
\be
\kappa_T^{-1} = (1+\eta)^2 \left[ 2f_2  
+ \sigma (\sigma - 1) f_{\sigma} |\eta|^{\sigma - 2} \right] =
-2 (\delta - 1) f_2(t) (1 \pm \eta_l)^2
\ee
where the upper sign is for the liquid side and the lower sign is for the gas side ($\eta_g = - \eta_l$).  This indicates that $f_2$ must of course be negative for $t < 0$.  The formula for the heat capacity along the coexistence curve is
\be
c_V = - \frac{1+t}{T_c} \left[ f^{\prime \prime}_0(t) + 
f^{\prime \prime}_1(t)\eta + f^{\prime \prime}_2(t)\eta^2 + 
f^{\prime \prime}_{\sigma}(t)|\eta|^{\sigma} \right] \, .
\ee
The singular part comes from the terms which are zero and second order in $\eta$.  This leads to $c_V \rightarrow c_- (-t)^{-\alpha}$ where 
\be
T_c \, c_- = (2-\alpha)(1-\alpha)a_- + \gamma (\gamma - 1) 
\left( \frac{2 \, b_-}{\sigma f_{\sigma}(0)} \right)^{\frac{2}{\delta - 1}} b_-
\, .
\ee 

According to Ref. \cite{Guida} the thermal compressibility $\kappa_T$ diverges as $\kappa_+ t^{-\gamma}$ when $t \rightarrow 0^+$ and as $\kappa_- (-t)^{-\gamma}$ when $t \rightarrow 0^-$, with $\kappa_+/\kappa_- \approx 5$ a universal ratio.  Also, the heat capacity at $\eta \rightarrow 0$ diverges as $c_+ t^{-\alpha}$ when $t \rightarrow 0^+$ and as $c_- (-t)^{-\alpha}$ when $t \rightarrow 0^-$, with $c_+/c_- \approx 0.5$ another universal ratio.  The former leads to the constraint
\be
b_+ = \frac{(\delta - 1) b_-}{5} 
\ee
while the latter leads to
\be
2 a_+ = a_- + \frac{\gamma (\gamma - 1)}{(2-\alpha)(1-\alpha)} 
\left( \frac{2 \, b_-}{\sigma f_{\sigma}(0)} \right)^{\frac{2}{\delta - 1}} b_-
\, .
\ee

If we are not too far from the critical point we can use the following parameterizations.  We can take $f_{\sigma}$ to be a constant.  The function
\ba
f_2(t) = \left\{ \begin{array}{ll}
\bar{f}_2(t)-b_- (-t)^{\gamma} & \mbox{if $t<0$}\\
\; \;\; \bar{f}_2(t) + b_+ t^{\gamma} & \mbox{if $t>0$}
\end{array} \right.
\ea
where $\bar{f}_2(t)$ is a smooth function which vanishes at $t=0$ as a power bigger than $\gamma$.  The function $f_1(t)$ is the chemical potential along the critical curve, which may be parameterized like this.  Assume a quadratic relationship between T and $\mu_x$.
\be
\left(\frac{T}{T_0}\right)^2 + \left(\frac{\mu_x}{\mu_0}\right)^2 = 1
\label{critcurve}
\ee  
The curve hits the $\mu$ axis at $T=0$ when $\mu=\mu_0$.  The chemical potential at the critical point is
\be
\mu_c = \mu_0 \sqrt{ 1 - \frac{T_c^2}{T_0^2} } \, ,
\ee
hence
\be
f_1(t) = n_c \mu_0 \sqrt{1 - \frac{T_c^2}{T_0^2}(1+t)^2 } \, .
\ee
It is apparent from the above expression for $f_1(t)$ that this whole parameterization is only good when $T < T_0$, otherwise $f_1(t)$ would become imaginary.

Knowing this, it is straightforward to derive a simple formula for the latent heat per unit volume.  Making use of phase equilibrium, $\Delta P = 0$, $T = T_x$, $\mu = \mu_x$, $\eta_g = - \eta_l$, one finds
\be
\Delta \epsilon(t) = \frac{2 n_c \mu_0^2 \eta_l(t)}{\mu_x(t)}
\ee
which of course is valid only for $t < 0$.

The independent constants may be taken as: $a_-, b_-, f_{\sigma}, \mu_0, \mu_c, T_c, n_c$, plus the functions $\bar{f}_0(t)$ and $\bar{f}_2(t)$.  This parameterization captures the critical exponents and amplitudes which are universal.  To proceed further we require more information on the equation of state of QCD away from the critical point.

For future reference it may be noted that one could add more terms to the free energy without changing the basic picture.  For example, one could add $f_4(t) \eta^4$ and $f_8(t) \eta^8$.  However, $f_4(t)$ must vanish at $t=0$ so as not to affect the critical behavior.  The coefficient $f_8(t)$ need not vanish at $t=0$ but it becomes irrelevant compared to the dominant term $f_{\sigma} |\eta|^{\sigma}$ as $\eta \rightarrow 0$, as do all powers of $\eta$ greater than $\sigma$.

\section{Fixing the Parameters}

In this section we narrow in on a phenomenological equation of state with input from various disparate sources.  These include the results of lattice gauge theory calculations at zero baryon density, and models and extrapolations of the equation of state of cold dense nuclear matter.

Concerning the function $\bar{f}_0(t)$, what we know from the thermodynamic relations is that $\bar{f}_0(0) = \mu_c n_c - P_c = \epsilon_c - T_c s_c$ and $\bar{f}^{\prime}_0(0) = - T_c s_c$.  Hence for small $t$ it starts out as $\bar{f}_0(t) = \epsilon_c - T_c s_c (1+t) + \cdot\cdot\cdot$.  

Suppose we wanted to extend this equation of state to $T=0$.  The minimum requirement is that the entropy density must vanish.  This means that $f^{\prime}_0(-1) = f^{\prime}_1(-1) = f^{\prime}_2(-1) = 
f^{\prime}_{\sigma}(-1) = 0$.  This is satisfied by the above parameterizations of $f_1$ and $f_{\sigma}$, but $f_0$ and $f_2$ must be modified.  The simplest way to make $f^{\prime}_2(-1)$ vanish which does not upset the critical point properties is to choose
\be
\bar{f}_2(t)=\thalf b_- \gamma t^2
\ee
following the idea that the function be smooth with the smallest integer power.  Note that with this choice $f_2(t)$ is negative for all $t<0$ and positive for all $t>0$.  The simplest way to make $f^{\prime}_0(-1)$ vanish, which does not upset the critical point properties, is to choose
\be
a_- = T_c s_c/(2-\alpha) \, .
\ee
If we then are so bold as to extrapolate our formula for the coexistence curve to $T=0$ we can input more physical information.

For example, suppose we know that phase coexistence at $T=0$ occurs between a liquid phase with density $n_l(T=0)$ and a gas phase with density $n_g(T=0)$.  Denoting normal nuclear density as $n_0$, we have $n_0 < n_g(T=0) < n_c < n_l(T=0)$ and $n_g(T=0) + n_l(T=0) = 2 n_c$.  In that case we can solve for $b_-$ in terms of the density difference $\Delta n = n_l(T=0) - n_g(T=0)$.
\be
b_- = \frac{\sigma f_{\sigma}}{2-\gamma} 
\left( \frac{\Delta n}{2n_c} \right)^{\delta - 1}
\ee
Then $\mu_0 = \mu(T=0)$ could be estimated by using the value calculated for nuclear matter at the density $n_g(T=0)$.  Now all the noncritical parameters would be determined apart from $f_{\sigma}$.

Consider some common parameterizations of the cold nuclear matter equation of state \cite{Grant}.  The energy density, pressure, and chemical potential are
\ba
\epsilon &=& n \left( m_N + E_0(n) \right) \\
P &=& n^2 \frac{dE_0(n)}{dn} \\
\mu &=& \frac{d\epsilon}{dn} \, .
\ea
Case I:
\ba
E_0(n) &=& \frac{K}{18} \left[ \frac{n}{n_0} - 1 \right]^2 + E_0(n_0) 
\nonumber \\
\mu(n) &=& m_N + E_0(n_0) + \frac{K}{18} \left[ \frac{n}{n_0} - 1 \right]
\left[ 3\frac{n}{n_0} - 1 \right]
\ea
Case II:
\ba
E_0(n) &=& \frac{2K}{9} \left[ (n/n_0)^{1/2} - 1 \right]^2 + E_0(n_0)
\nonumber \\
\mu(n) &=& m_N + E_0(n_0) + \frac{2K}{9} \left[ (n/n_0)^{1/2} - 1 \right]
\left[ 2(n/n_0)^{1/2} - 1 \right]
\ea
Here $m_N = 939$ MeV is the nucleon mass and $E_0(n_0) = - 16.3$ MeV is the average binding energy per nucleon at nuclear matter density $n_0 = 0.153/{\rm fm}^3$ \cite{nuc1}.  The compressibility $K$ is known to be $250 \pm 30$ MeV \cite{nuc1,nuc2}; we shall fix it at 250.  See also \cite{KapGale}.

Heavy ion collisions at the Bevalac and at the AGS showed no clear experimental evidence for the formation of quark-gluon plasma \cite{HIreviews}.  The baryon densities achieved were around two to four times nuclear matter density.  If one distributes one unit of baryon number within one electromagnetic radius of a proton, 0.8 fm, the baryon density would be about 0.47/fm$^3$, which is slightly more than three times nuclear matter density.  Therefore, it seems reasonable to estimate $n_g(T=0) = 4 n_0$.  At this density case I gives $\epsilon(4n_0) = 641$ MeV/fm$^3$ and $\mu(4n_0) = 1381$ MeV.  Case II gives $\epsilon(4n_0) = 599$ MeV/fm$^3$ and $\mu(4n_0) = 1089$ MeV.  Case I corresponds to a relatively stiff equation of state whose energy per baryon rises quadratically at high density, whereas case II corresponds to a relatively soft equation of state whose energy per baryon rises linearly at high density.  The energy densities are similar because they are dominated by the nucleon mass, not interactions.  The chemical potentials differ by about 20\% because interactions do contribute.  The pressure is most sensitive to the interactions.  Based on this information we estimate $\mu_0 = 1230 \pm 150$ MeV.

The Hagedorn temperature was already determined in the late 1960's and early 1970's to be 160 MeV \cite{Hagedorn}.  The critical temperature, no matter what order the transition is, ought to be slightly greater than this \cite{excluded}.  Data from heavy ion collisions at the SPS and RHIC show that no hadrons have ever been observed with a temperature greater than about 160 to 170 MeV (at very small chemical potential) \cite{statmodels}.  Current lattice QCD calculations agree that with the physical values of the light and strange quark masses, the transition is a rapid crossover at zero chemical potential.  However, they disagree on the so-called critical temperature.  One group \cite{Fodor} puts it at 150 MeV while the other group \cite{HotQCD} puts it at 190 MeV.  Part of the discrepancy may be in exactly how this temperature is defined, but that is not entirely sufficient.  Certainly, to accurately determine this temperature requires an accurate calculation of the low temperature hadronic equation of state.  But this requires very fine lattice spacing, since the lattice must first discern the structure of individual hadrons, and it requires a very large lattice volume, since the hadrons become widely separated at low temperature.  This is a difficult problem which may not be resolved for some time.  However, it seems safe to estimate $T_0 = 170 \pm 20$ MeV.

The value of the temperature at the critical point is of course not known.  However, it should lie on or very near to the curve of $T$ versus $\mu$ under discussion.  How then can we estimate the pressure $P_c$, energy density $\epsilon_c$, entropy density $s_c$, and baryon density $n_c$ at the critical point?  One obvious way is to use the formulas for a perfect massless gas of gluons and $N_f$ flavors of quarks evaluated at $T_c$ and $\mu_c$.  These formulas are ($\mu$ is the baryon chemical potential and quarks have one third of that value):    
\ba
P &=& \frac{\pi^2}{90} \left( 16 + \frac{21N_f}{2} \right) T^4 +
\frac{N_f}{18} \mu^2 T^2 + \frac{N_f}{324\pi^2} \mu^4 \\
s &=& \frac{4\pi^2}{90} \left( 16 + \frac{21N_f}{2} \right) T^3 +
\frac{N_f}{9} \mu^2 T \\ 
n &=& \frac{N_f}{9} \mu T^2 + \frac{N_f}{81\pi^2} \mu^3 \\
\epsilon &=& 3P \, .
\ea
When the relationship
\be
\mu_c^2 = \mu_0^2 \left( 1 - \frac{T_c^2}{T_0^2} \right)
\ee
with the aforementioned estimates of $T_0$ and $\mu_0$ is used, it turns out that the pressure is almost independent of the numerical value of $T_c$.  Since two phases in equilibrium with the same $T$ and $\mu$ have the same pressure, wouldn't it be nice if $P_c$ was independent of $T_c$?  This is one hint.  A second hint is provided by the fact that all lattice QCD calculations show that the pressure, energy density, and entropy density are all lower than the ideal gas formula would suggest, at least at $\mu = 0$.  In fact, they indicate a negative contribution to the pressure proportional to $T^2$ \cite{Boyd}-\cite{Tsquare}, and a negative constant contribution, like a bag constant.  It has also been suggested, in the context of cold dense matter as might exist in neutron stars, that at $T=0$ there may be a contribution proportional to $\mu^2$ \cite{musquare}.  So let us hypothesize that, in the vicinity of the phase transition or crossover and above, the high energy density equation of state can be parameterized as
\be
P = A_4 T^4 + A_2 \mu^2 T^2 + A_0 \mu^4 - C T^2 - D \mu^2 - B
\ee
where $A_4$, $A_2$ and $A_0$ are given by the perfect gas equation of state.  Now suppose we substitute $\mu^2 = \mu_0^2 (1-T^2/T_0^2)$ into this formula, and demand that it be independent of $T$.  A simple exercise shows that $T_0$ and $\mu_0$ must be related according to
\be
\frac{\mu_0^2}{T_0^2} = 9\pi^2 \left[ 1 \pm
\sqrt{ \frac{8}{15} - \frac{32}{45 N_f} } \right] \, .
\ee
Choosing $N_f = 2.5$, effectively to account for the smaller contribution from the heavier strange quarks at these temperatures and chemical potentials, and choosing the minus sign, leads to
\be
\frac{\mu_0}{T_0} = 6.67173...
\ee
in order that the coefficient of $T^4$ vanish.  In other words, if $T_0=180$ MeV then $\mu_0 = 1209$ MeV.  This relationship is entirely in line with all the facts at hand.  Demanding that the coefficient of the $T^2$ term vanish requires
\be
C - \frac{\mu_0^2}{T_0^2} D = \mu_0^2 \left[ A_2 - 2 \frac{\mu_0^2}{T_0^2} A_0
\right] \approx 3.084 T_0^2 \, .
\ee
The lattice calculations of \cite{HotQCD} found that $2C \approx 0.24$ GeV$^2$.  This translates into $C \approx 3.3 T_0^2$ using their value of $T_0 \approx 190$ MeV.  There are no calculations of the $\mu^2$ term in the pressure, but this analysis suggests that $D$ is very small; we shall take it to be zero for simplicity of exposition.  

With even larger uncertainties Ref. \cite{HotQCD} found that $B \sim T_0^4$.  For a reasonable interpolation of the lattice results near and just above the crossover region we take the coefficient to be 0.8.  The parameterization is therefore
\bd
P = \frac{169\pi^2}{360} T^4 +
\frac{5}{36} \mu^2 T^2 + \frac{5}{648\pi^2} \mu^4
- 3.084 T_0^2 T^2 - 0.8 T_0^4
\ed
\bd
s = \frac{169\pi^2}{90} T^3 + \frac{5}{18} \mu^2 T - 6.168 T_0^2 T
\ed
\bd
n = \frac{5}{18} \mu T^2 + \frac{5}{162\pi^2} \mu^3
\ed
\be
\epsilon = - P +Ts + \mu n \, .
\label{QGPpara}
\ee
The critical pressure is computed from this to be $P_c = 0.749 T_0^4$.  The critical entropy density $s_c$, baryon density $n_c$, and energy density $\epsilon_c$ of course depend on the choice of $T_c$ and therefore $\mu_c$.

All that remains is to specify $\Delta n$ at $T=0$ and $f_{\sigma}$.  Since $f_{\sigma}$ is assumed to be constant, it is natural that it be proportional to $P_c$.  For definiteness we shall take $f_{\sigma} = 5P_c \approx 512$ MeV/fm$^3$ and $\Delta n = n_c/3$.  In what follows we shall use all of the above parameterizations and only vary $T_c$ to see what effect it might have on heavy ion collisions.

\section{Numerical Results for Equation of State}

In this section we plot some of the thermodynamic functions that were derived in the previous two sections.  It is important to note that, although the results do depend on the numerical values of the parameters, the critical behaviors obviously do not.  In addition, since most of the parameters were chosen to match onto known properties of quark-gluon matter at $\mu = 0$ and to dense nuclear matter at $T = 0$ the results should not be too far from what is at present impossibly difficult computations in QCD.

\newpage

In Fig. 1 we show the pressure, entropy density, and energy density obtained from eq. (\ref{QGPpara}) as functions of $T$ at $\mu = 0$.  They are close to the curves computed in lattice QCD but not identical.  The parameterization of eq. (\ref{QGPpara}) is needed to extrapolate the lattice results to large chemical potentials.  Anyway, all that is needed for the purposes of the chiral critical point are the values of $P_c$, $s_c$ and $\epsilon_c$ for a chosen value of $T_c$, not the full $T$ and $\mu$ dependence of the equation of state of quark-gluon plasma.

\begin{figure}[hb]
\begin{center}\includegraphics[width=4.3in,angle=90]{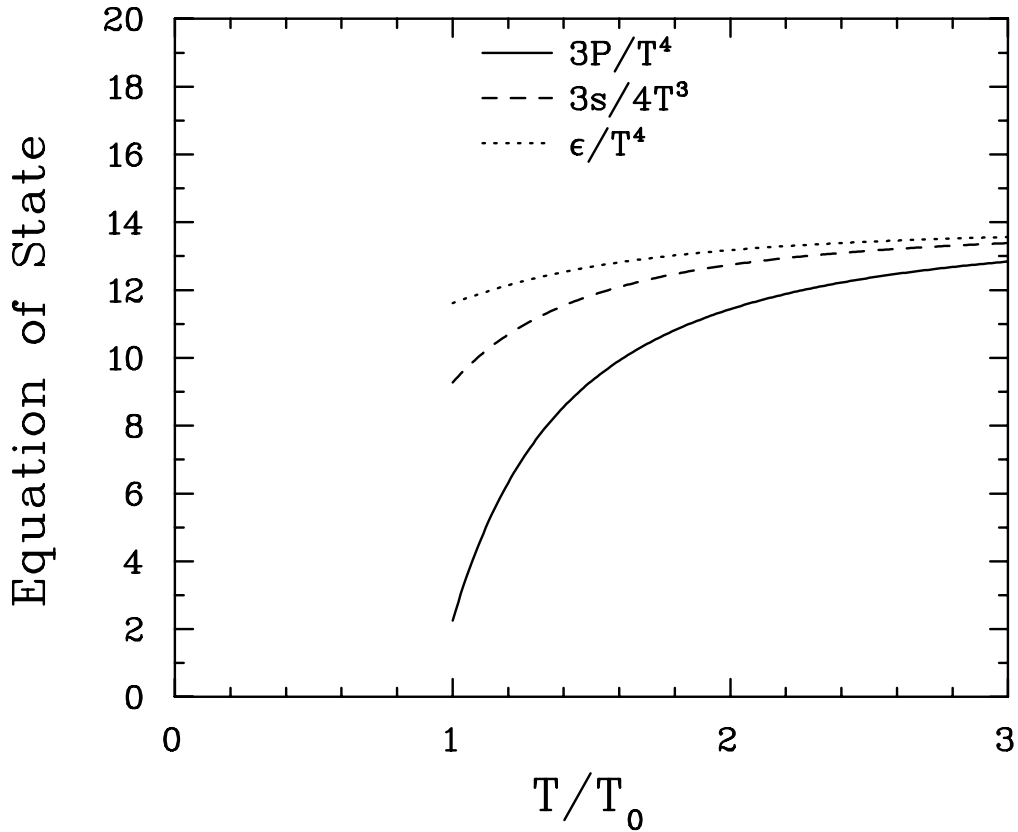}
\caption{The pressure, entropy density, and energy density, normalized so that they all have the same asymptotic value, versus temperature at $\mu=0$.  The parameterization is from eq. (\ref{QGPpara}) which is motivated by lattice QCD calculations.}
\end{center}
\label{eos4qgp}
\end{figure}

\newpage

In Fig. 2 we show the curve of phase coexistence, $T$ versus $\mu$, according to eq. (\ref{critcurve}).  One chooses $T_c$ somewhere along this curve.  Then for $T > T_c$ along this curve the transition is a rapid crossover, whereas for $T < T_c$ along this curve the transition is first order.

\begin{figure}[hb]

\begin{center}\includegraphics[width=4.2in,angle=90]{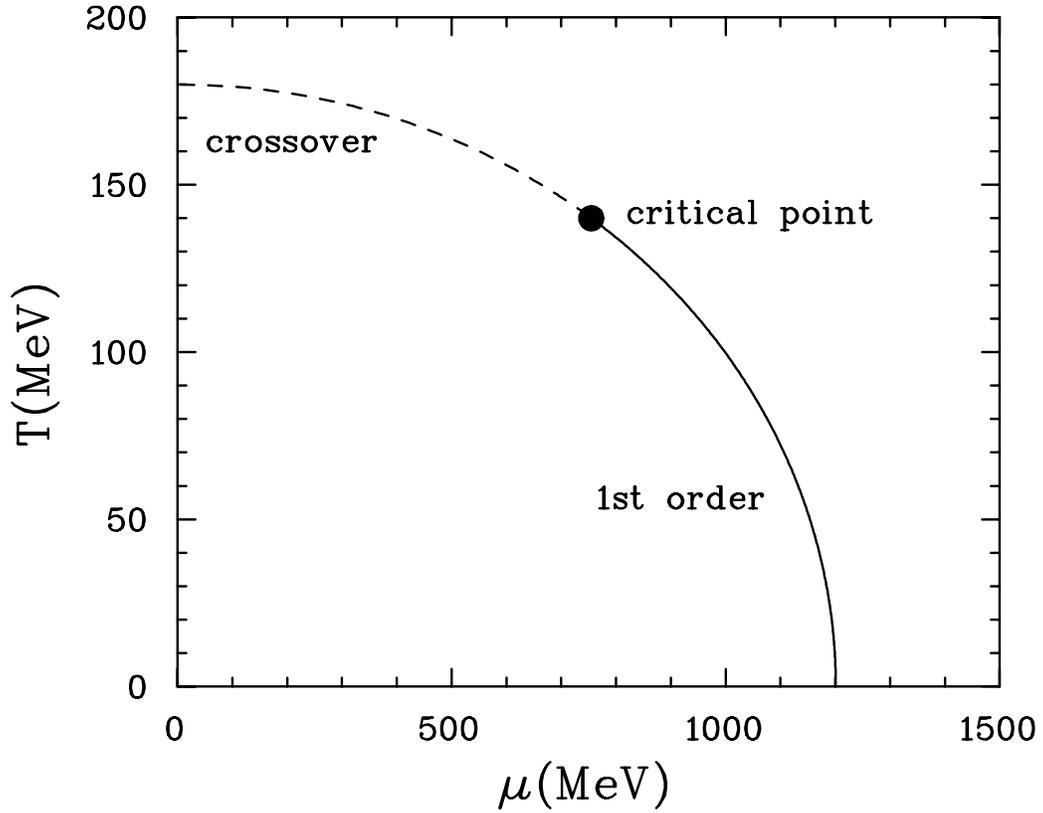}
\caption{Temperature versus baryon chemical potential from the parameterization of eq. (\ref{critcurve}).  The critical temperature lies somewhere along this curve.}
\end{center}
\label{Tvsmu}
\end{figure}

\newpage

In Fig. 3 we show the curve of phase coexistence, $T$ versus $n/n_0$, for various choices of $T_c$.  These are indicated by the solid curve.  The dashed curve indicates the limit of isothermal metastability, or isothermal spinodal.  When $T$ is scaled by $T_c$ the phase coexistence curves fall on top of one another, as do the spinodals, indicative of a special scaling feature of this parameterization.

\begin{figure}[hb]
\begin{center}
\includegraphics[width=0.4\textwidth,angle=90]{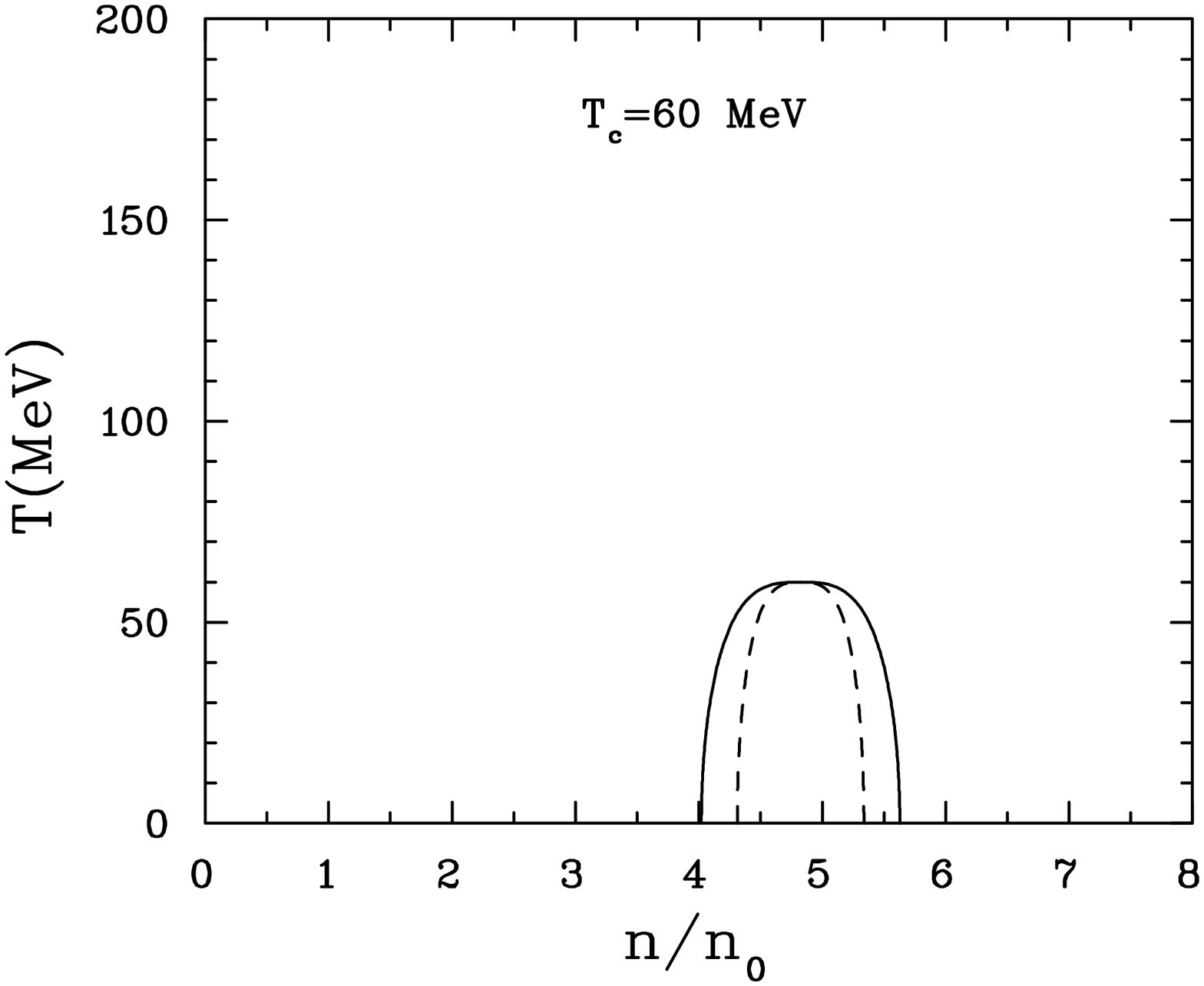}
\includegraphics[width=0.4\textwidth,angle=90]{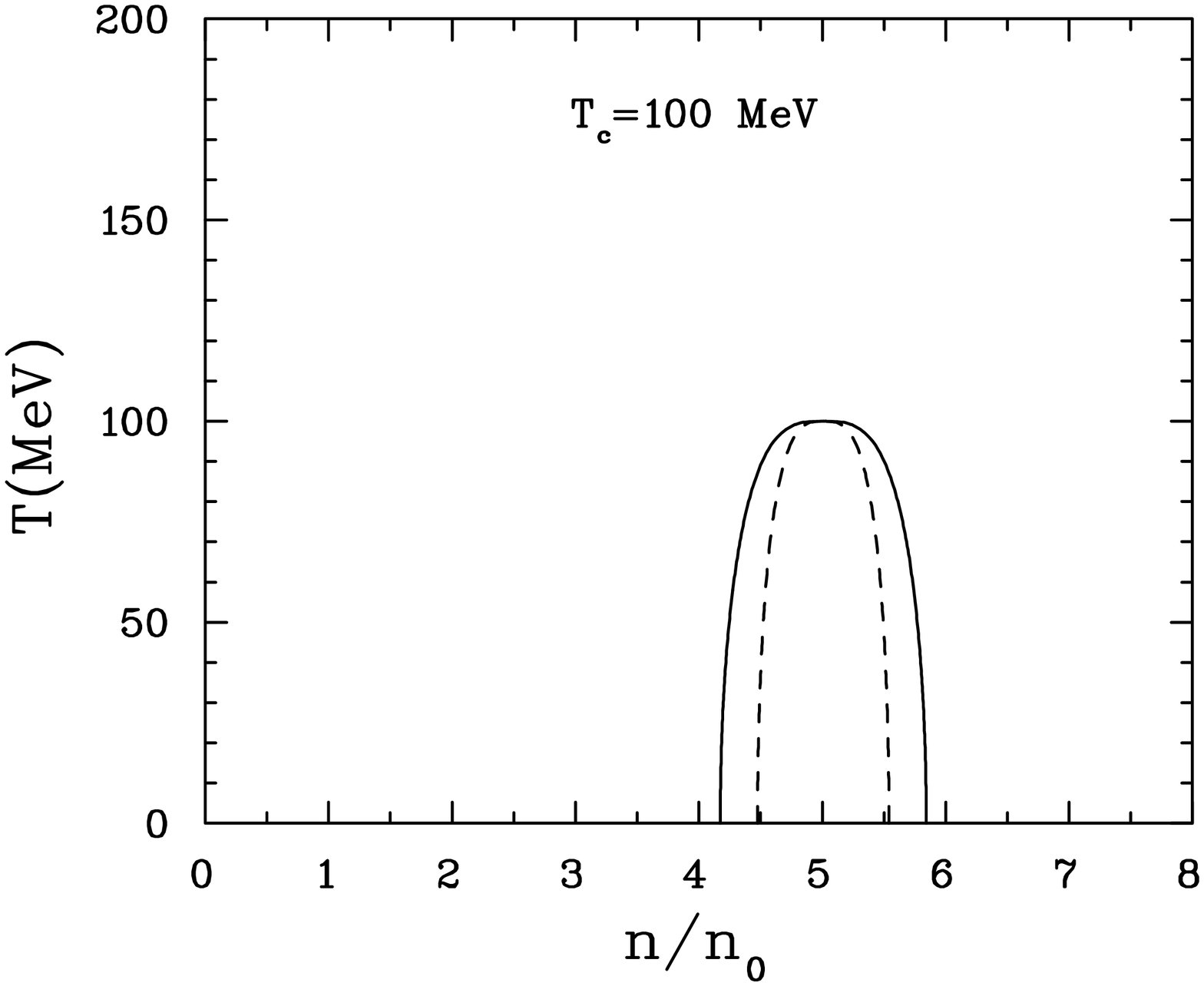}
\includegraphics[width=0.4\textwidth,angle=90]{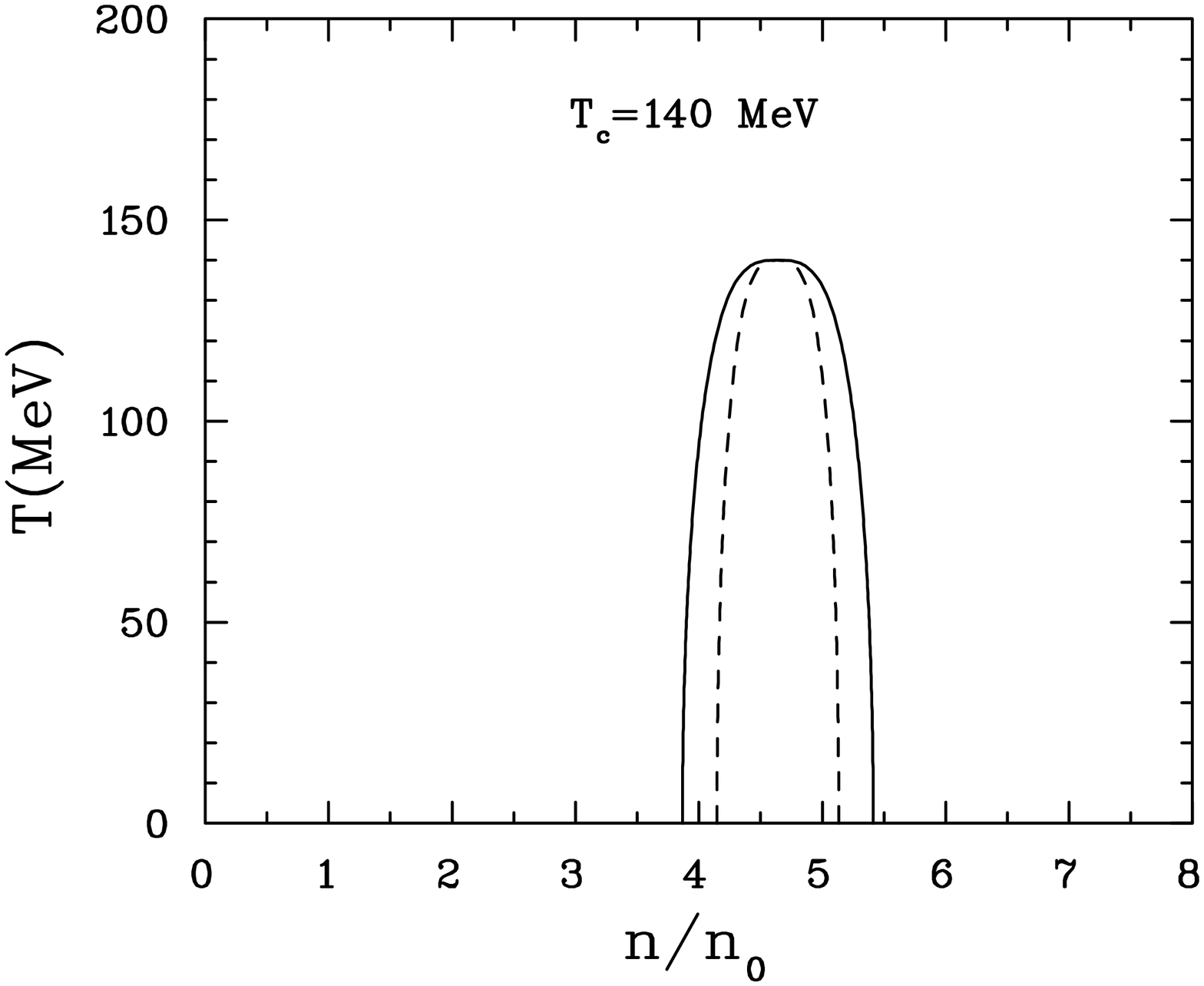}
\includegraphics[width=0.4\textwidth,angle=90]{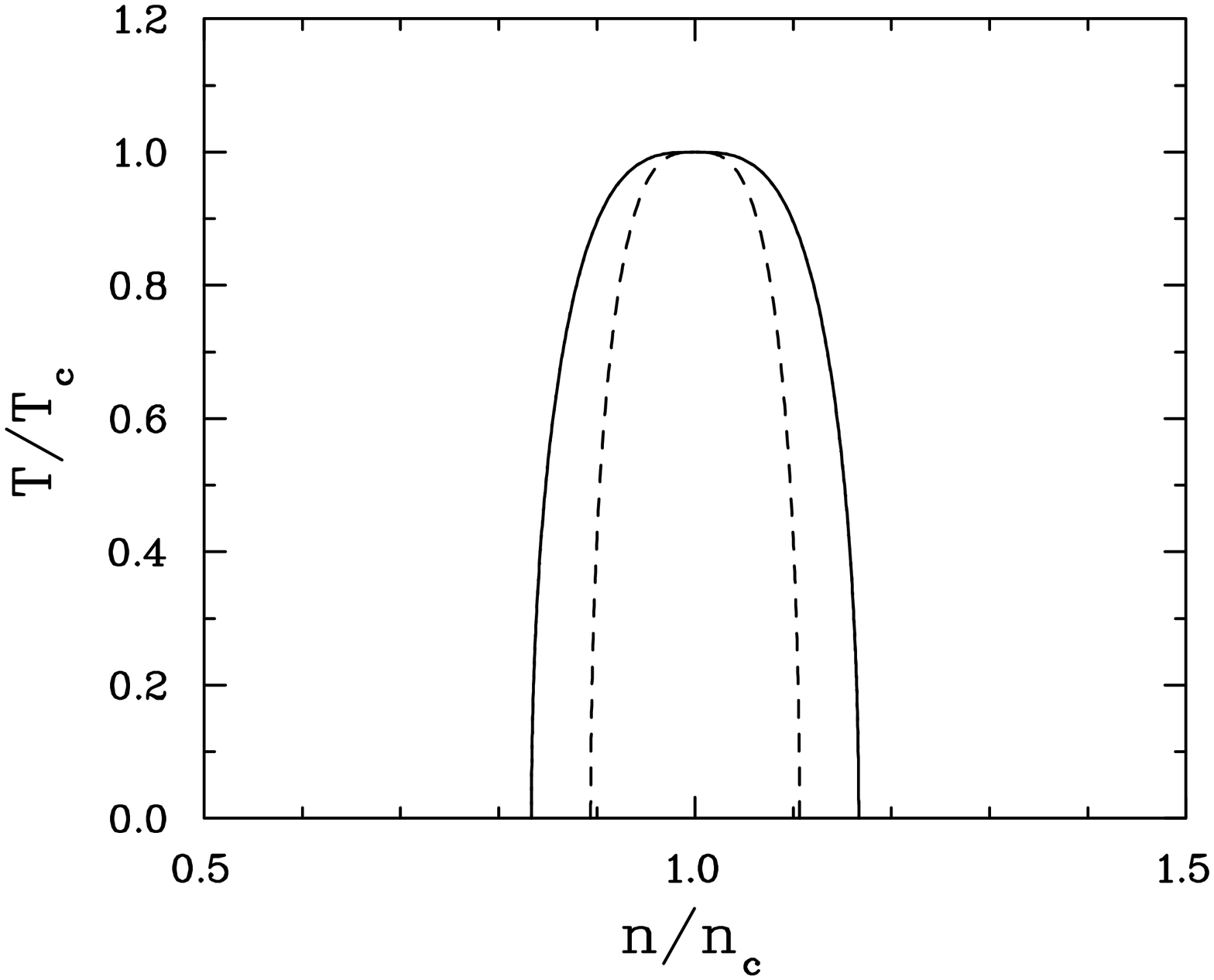}
\caption{The solid curve denotes coexistence between high and low density phases.  The dashed curve denotes the limits of metastability.  When scaled by the critical temperature and density the curves lie on top of each other.}
\end{center}
\label{Tvsn}
\end{figure}

\newpage

In Fig. 4 we show the isothermal compressibility as a function of the temperature.  For $t < 0$ it depends on whether one approaches the phase coexistence curve from the low density side or the high density side.  For $t > 0$ it is computed at the critical density.  The result is independent of the choice of $T_c$ in this parameterization of the equation of state.

\begin{figure}[hb]
\begin{center}\includegraphics[width=4.3in,angle=90]{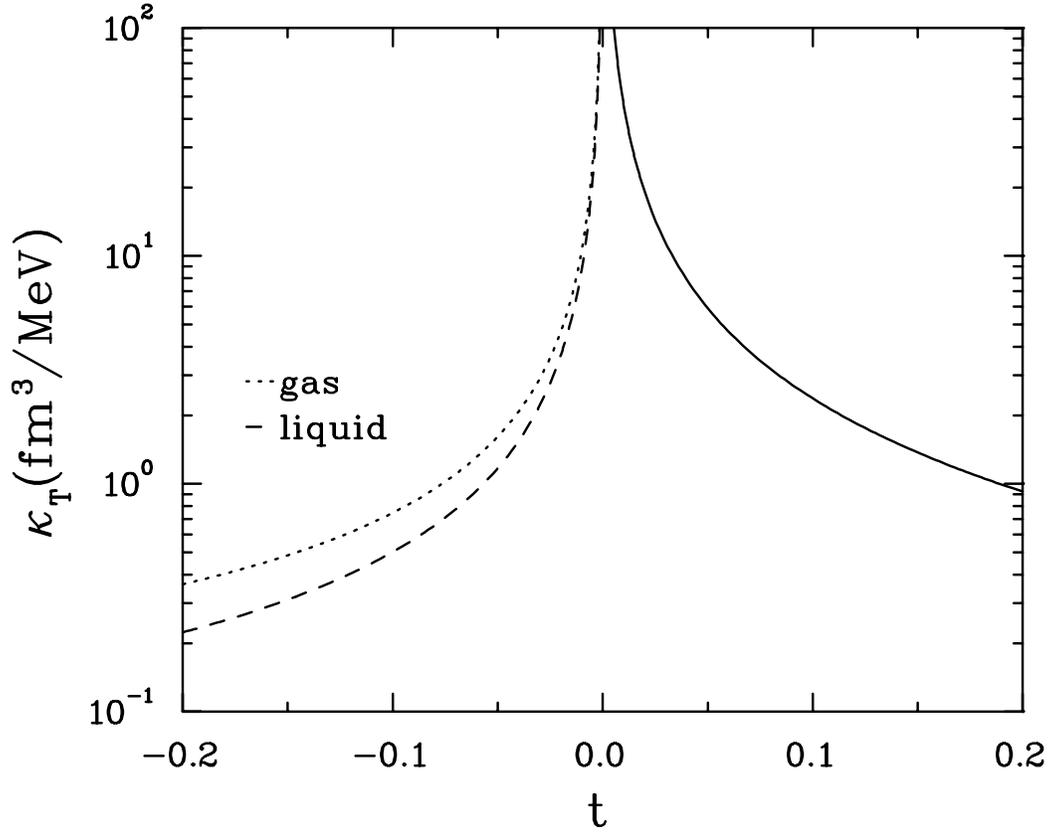}
\caption{The isothermal compressibility.  For $t<0$ they are evaluated along the coexistence curve while for $t>0$ it is evaluated at the critical density.  The curves for $T_c$ = 60, 100, and 140 MeV lie on top of one another.}
\end{center}
\label{kappa}
\end{figure}

\newpage

In Fig. 5 we show the heat capacity per unit volume as a function of temperature for various choices of $T_c$. For $t < 0$ it depends on whether one approaches the phase coexistence curve from the low density side or the high density side.  For $t > 0$ it is computed at the critical density.  When scaled by the entropy density at the critical point, the result for $t > 0$ is independent of the choice of $T_c$, whereas for $t < 0$ it is almost but not quite independent.  For reference, the entropy density at the critical point is 1.741, 3.416, and 5.861 fm$^{-3}$ for $T_c$ = 60, 100, and 140 MeV, respectively.

\begin{figure}[hb]
\begin{center}
\includegraphics[width=0.39\textwidth,angle=90]{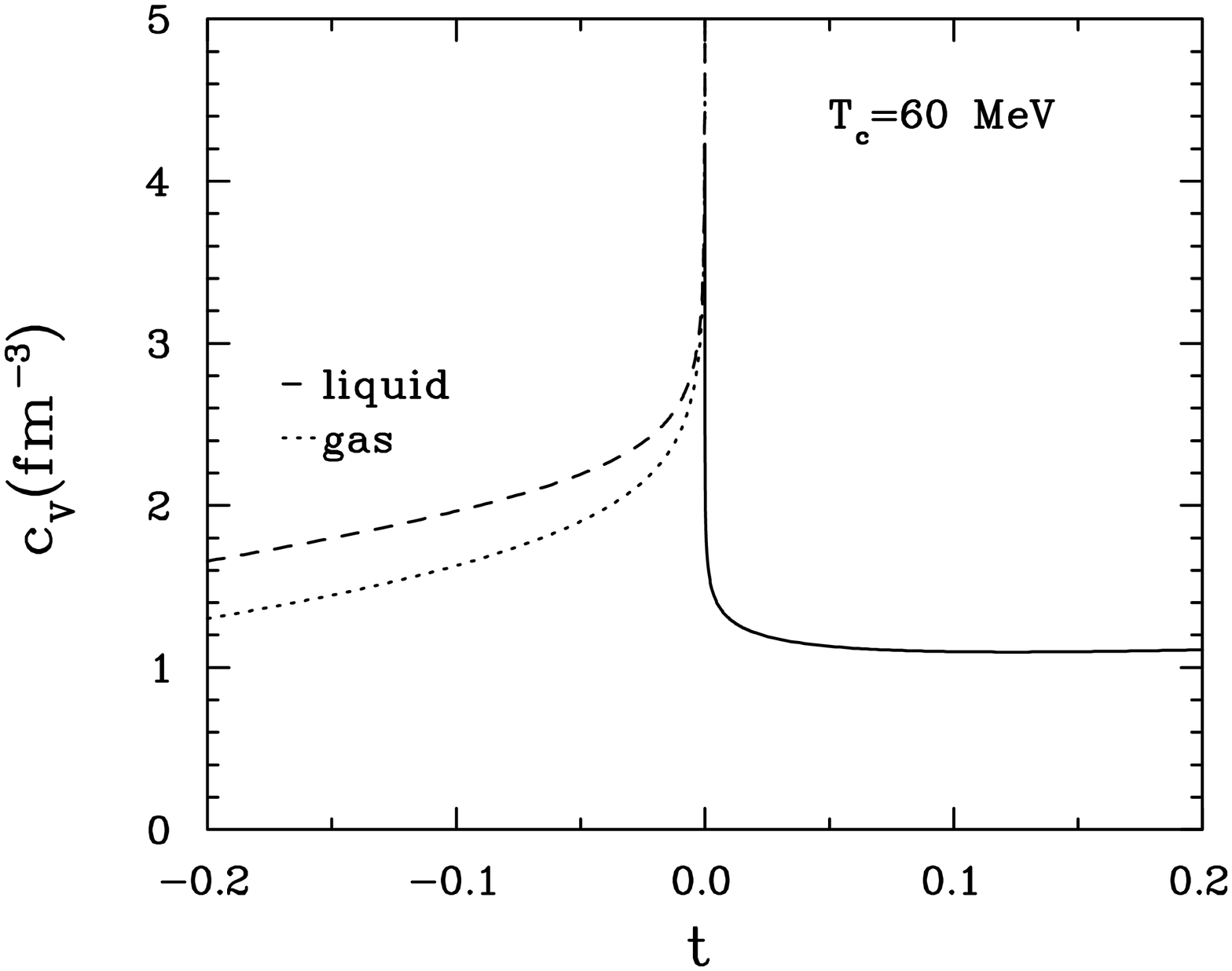}
\includegraphics[width=0.39\textwidth,angle=90]{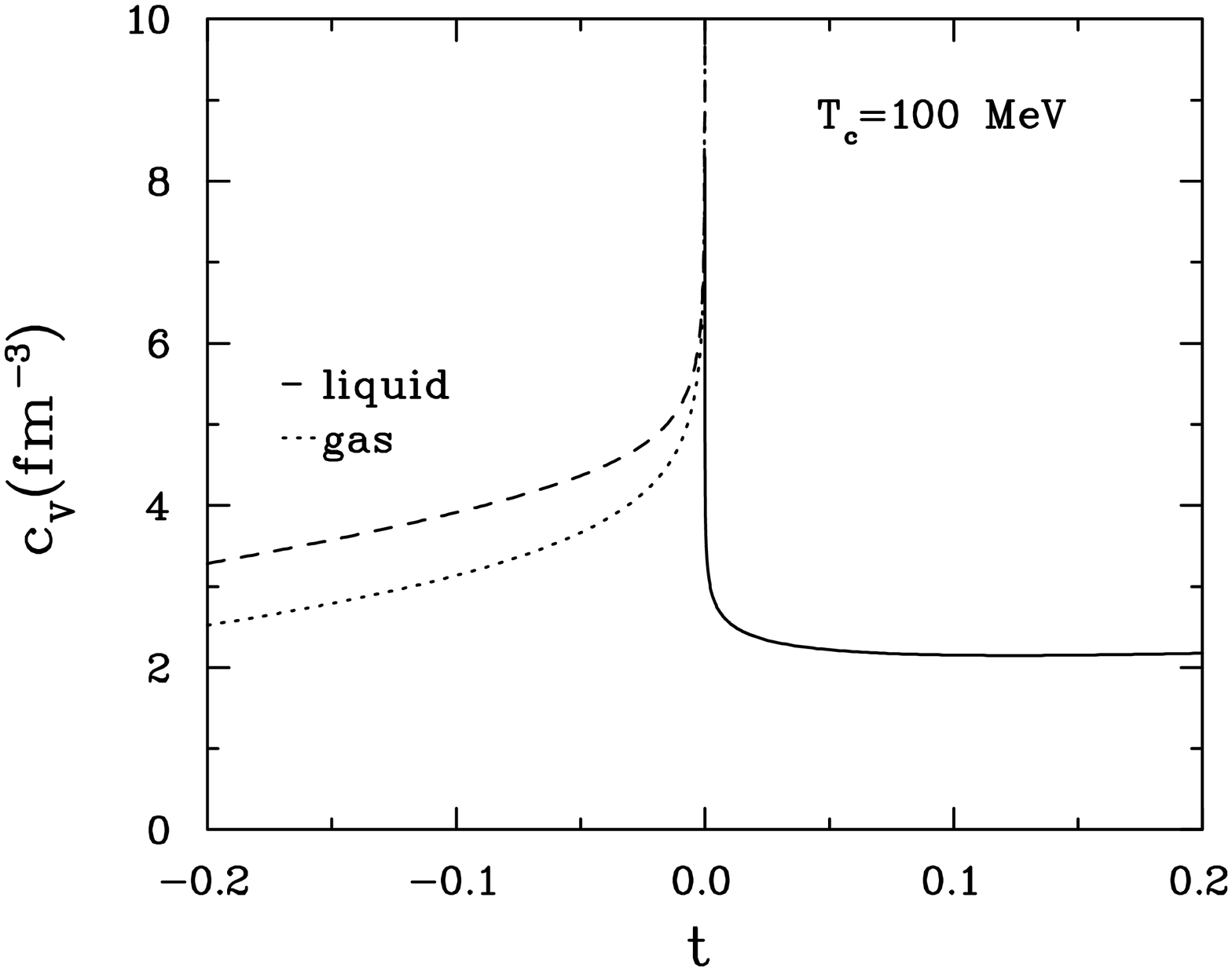}
\includegraphics[width=0.39\textwidth,angle=90]{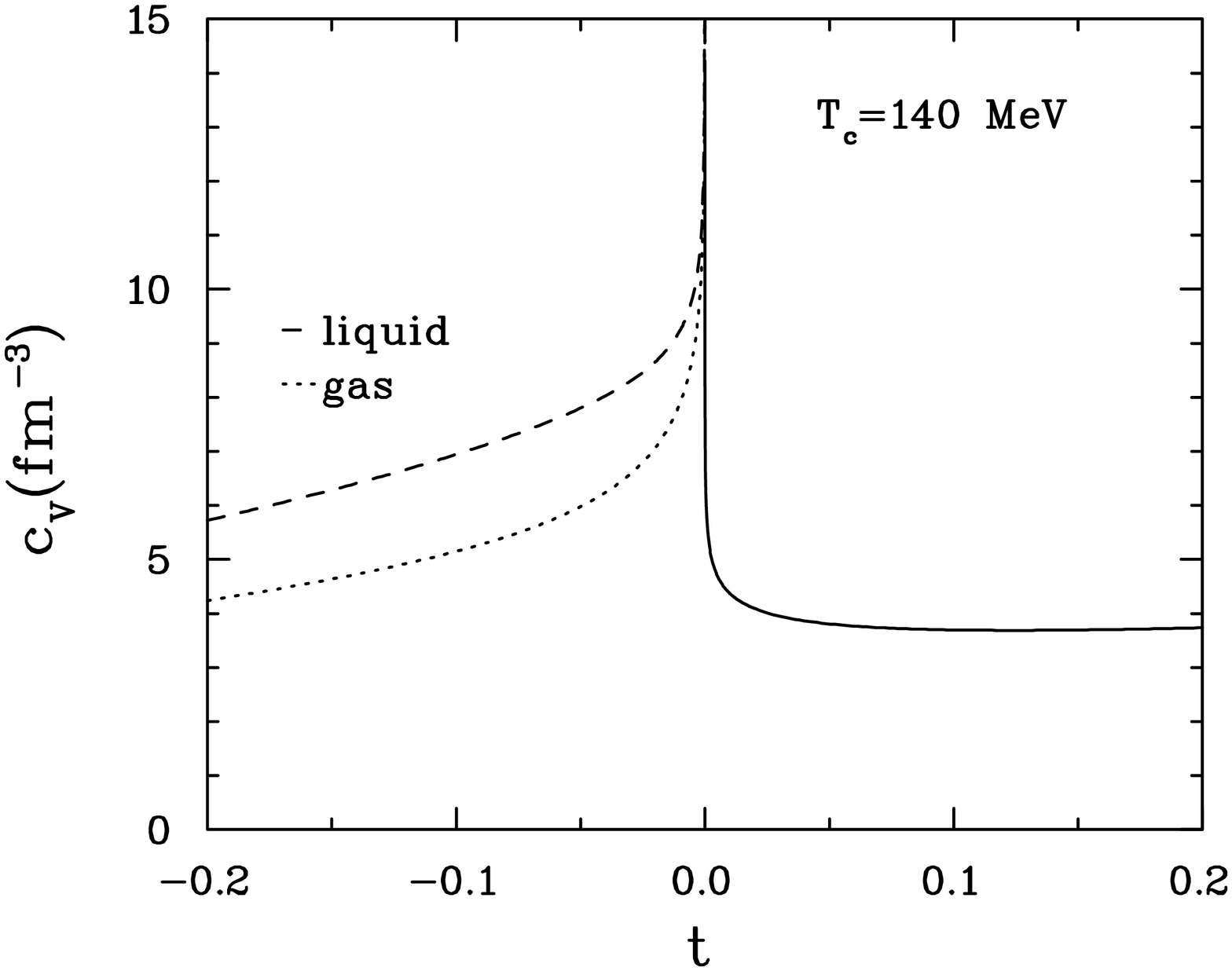}
\includegraphics[width=0.39\textwidth,angle=90]{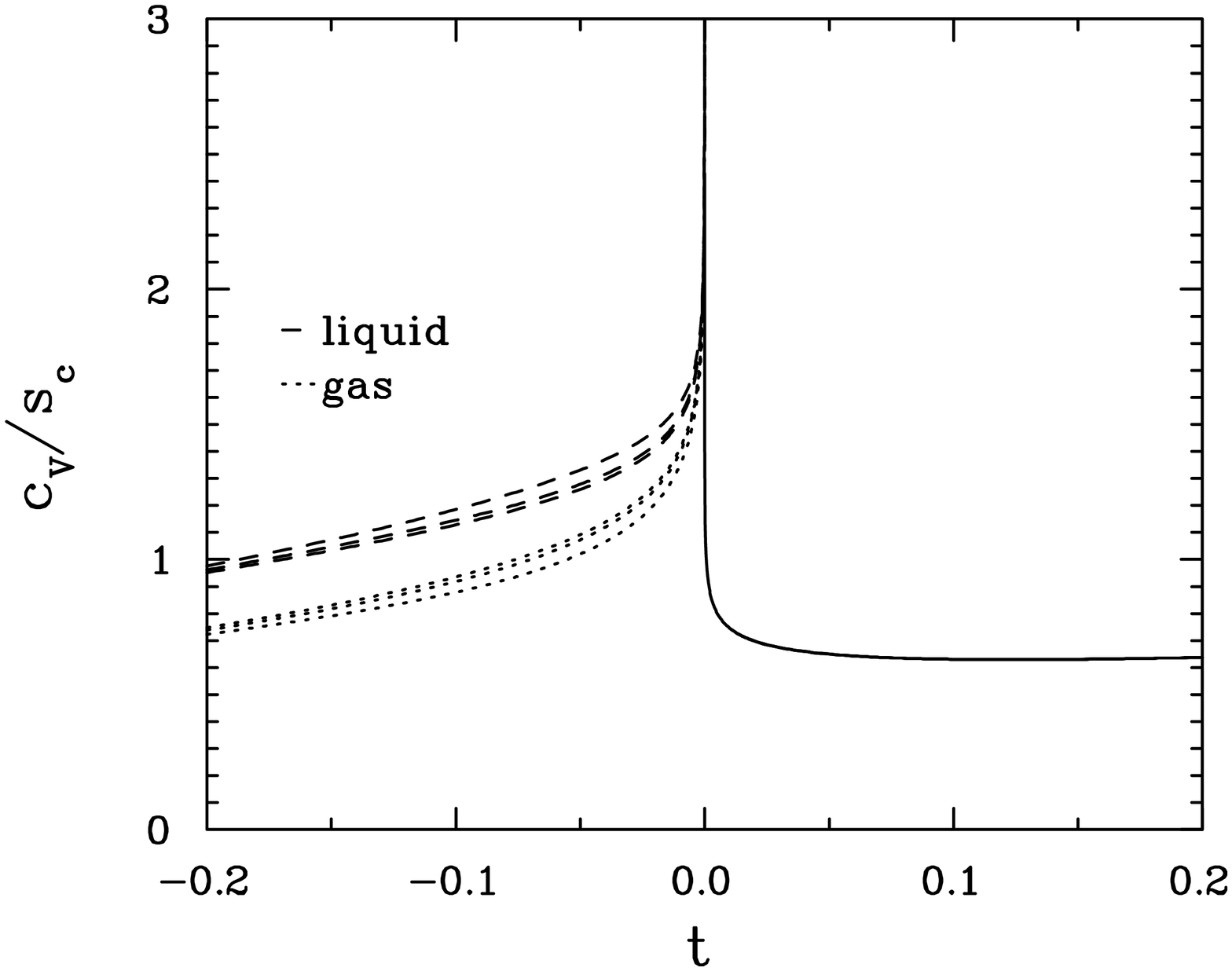}
\caption{ The heat capacity per unit volume.  For $t<0$ they are evaluated along the coexistence curve while for $t>0$ it is evaluated at the critical density.  When divided by the entropy density at the critical point the results are nearly independent of $T_c$.}
\end{center}
\label{cv}
\end{figure}

\newpage

In Fig. 6 we show the latent heat, or discontinuity in energy density, as a function of $T$ for various values of $T_c$.  For $60 < T_c < 140$ MeV the latent heat is approximately 300 MeV/fm$^{3}$ at $T=0$ and goes to zero at $T_c$. 

\begin{figure}[hb]
\begin{center}\includegraphics[width=4.2in,angle=90]{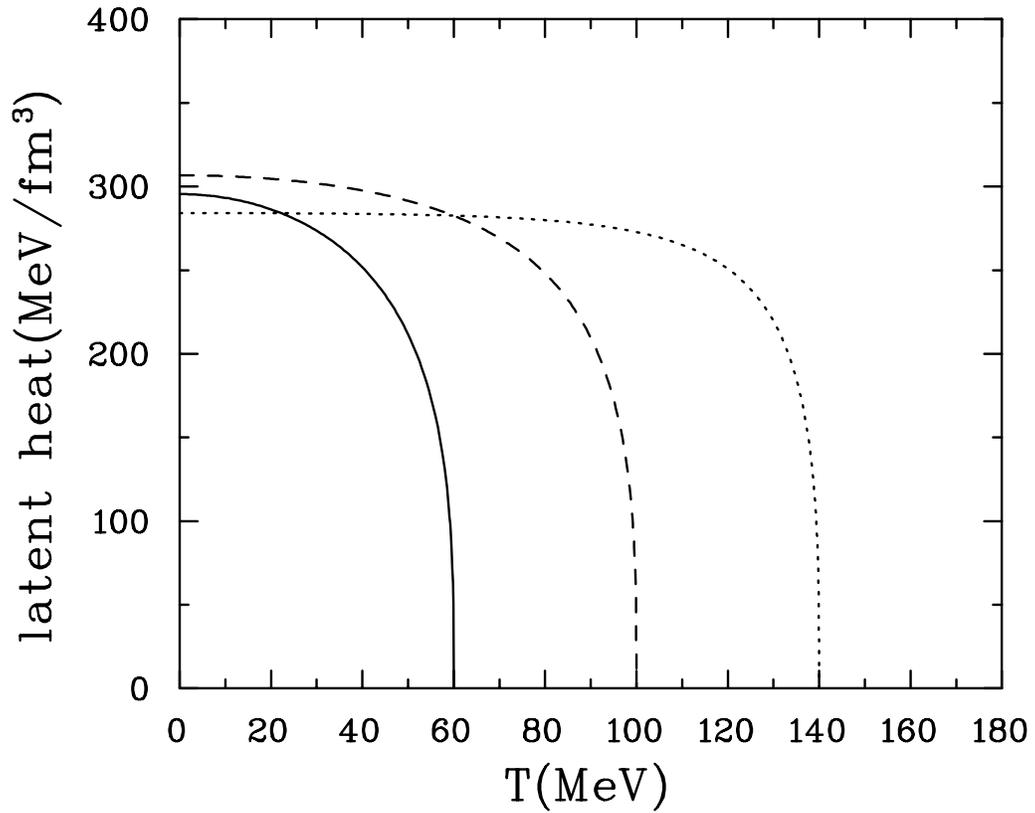}
\caption{The latent heat per unit volume versus temperature for three choices of $T_c$.}
\end{center}
\label{latent}
\end{figure}

\newpage

\section{Fluctuations}

All thermodynamic functions are smooth and continuous for any finite volume.  The discontinuities associated with phase transitions only arise in the infinite volume limit.  The question to be addressed here is whether the features characteristic of a chiral phase transition get smoothed out in heavy ion collisions to such an extent that they cannot be discerned from experimental observations.  The analysis performed here is based on the Landau theory of fluctuations \cite{Landau}, suitably modified to take into account the 
non-integer powers of $\eta$ appearing in the thermodynamic functions.

In a uniform system of large but finite volume $V$, the thermodynamic potential $\Omega$ depends only on the temperature and baryon chemical potential and is proportional to $V$.  Due to finite size fluctuations at the given temperature and chemical potential, the actual value of the ``order parameter'' $\eta$ will not necessarily be the equilibrium one.  To quantify this phenomenon we expand $\Omega$ in powers of $\eta$.
\be
\Omega(\mu,T;\eta) = \Omega_0(\mu,T) + \Omega_1(\mu,T)\eta +\Omega_2(\mu,T)\eta^2 +\Omega_{\sigma}(\mu,T)|\eta|^{\sigma}
\ee
In equilibrium this must be an extremum with respect of variations in $\eta$, namely
\be
\frac{\partial \Omega(\mu,T;\eta)}{\partial \eta} = 
\Omega_1(\mu,T) + 2\Omega_2(\mu,T)\eta + \sigma \Omega_{\sigma}(\mu,T)
|\eta|^{\sigma - 1} \, {\rm sign}(\eta) = 0 \, .   
\ee
The coefficient functions are determined by the fact that this condition is fulfilled by the equation of state.  Noting the powers of $\eta$ which appear, it is clear that one should choose
\be
\Omega_1(\mu,T) = K (-n_c \mu + f_1)
\ee
where $K$ is some factor yet to be fixed.  Upon using the equilibrium relation between $\mu$ and $\eta$, namely eq. (\ref{chempot}), one can determine the other coefficient functions.
\ba
\Omega_2 &=& K f_2 \nonumber \\ 
\Omega_{\sigma} &=& K f_{\sigma}
\ea
Therefore
\be
\Omega(\mu,T;\eta) = \Omega_0(\mu,T) + K \left[ (-n_c \mu + f_1) \eta
+ f_2 \eta^2 + f_{\sigma}|\eta|^{\sigma} \right] \, .
\ee
In equilibrium $\Omega/V = - P$.  Comparing this with the expression (\ref{pressure}) for the pressure, one can deduce that $K=V$ and
\be
\Omega_0(\mu,T) = V \left( f_0 - n_c \mu \right) \, .
\ee
Hence we have obtained the expansion around the equilibrium states
\be
\Omega(\mu,T;\eta) = \Omega_0(\mu,T) + V \left[ (-n_c \mu + f_1) \eta
+ f_2 \eta^2 + f_{\sigma}|\eta|^{\sigma} \right]
\ee 
with $\Omega_0(\mu,T)$ given above.

The probability ${\cal P}(\eta)$ to find the system with a particular value of $\eta$ at given values of $\mu$ and $T$ is
\be
{\cal P}(\eta) \sim {\rm e}^{-\Omega(\mu,T;\eta)/T} \, .
\ee
Along the coexistence curve $n_c \mu = f_1(t)$.  Then
\be
\Omega(\mu,T;\eta) - \Omega_0(\mu,T) = V \left[
f_2 \eta^2 + f_{\sigma}|\eta|^{\sigma} \right] \, .
\ee 
Since $t<0$, $f_2(t)<0$ and the thermodynamic potential has two equal minima at the densities of the liquid and gas phases, of course.  This potential is shown in Fig. 7 for temperatures both below and above $T_c$.  In this figure the volume was taken to be 400 fm$^3$.  Obviously the potential scales proportionately with this volume.  The value of 400 fm$^3$ is really quite optimistic for high energy nuclear collisions.  Considering that the critical density is estimated to be about $5n_0 \approx 0.75$ baryons/fm$^3$, this would mean that about 300 baryons participate in the fluctuation.  That is a substantial fraction of the total of 394 in Au+AU, 416 in Pb+Pb, and 476 in U+U collisions.  Even then, the potential for the low and high density phases are only 5 MeV below the unstable mid-point when $T/T_c = 0.6$; it is even less as $T_c$ is approached. 

\begin{figure}
\begin{center}\includegraphics[width=4.3in,angle=90]{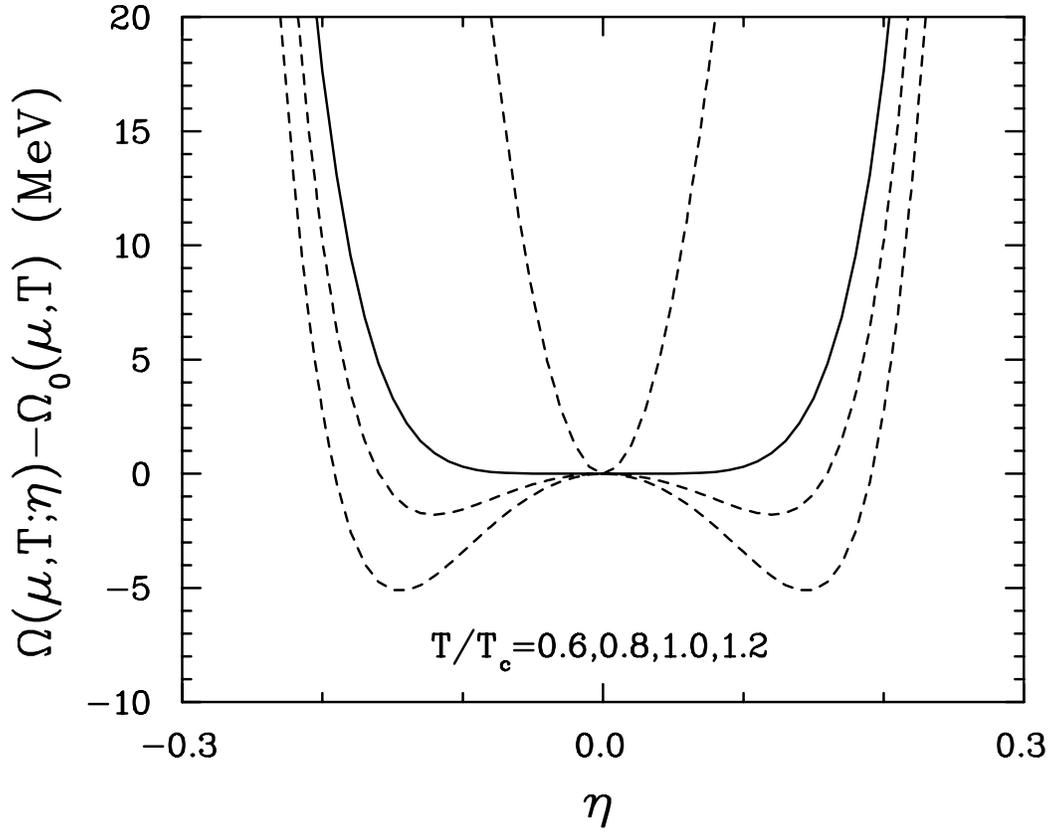}
\caption{The thermodynamic potential as a function of $\eta$ near the critical point; the volume is 400 fm$^3$.  The stable phases are located at the minima of the potential.  Four different temperatures are shown, with the solid curve representing the critical temperature.}
\end{center}
\label{potdif}
\end{figure}

It is interesting to find the probability that the system has some value of $\eta$ other than $\eta_g$ or $\eta_l$ along the curve of phase coexistence.  The relative probability is
\be
{\cal P}(\eta)/{\cal P}(\eta_l) =
 {\rm e}^{-\Delta \Omega/T}
\ee
where
\ba
\Delta \Omega &=& \Omega(\mu_x(T),T;\eta) - \Omega(\mu_x(T),T;\eta_l) 
\nonumber \\
&=& V \left[ f_2 \left( \eta^2 - \eta_l^2 \right)
+ f_{\sigma} \left( |\eta|^{\sigma} - |\eta_l|^{\sigma} \right) \right] \, .
\ea
Still using $V = 400$ fm$^3$, the relative probability is plotted as a function of $\eta$ in Fig. 8 for several values of $T \le T_c$. For $T \ge 0.6T_c$ there is more than a 90\% probability to find the system with any value of $\eta$ in the range from -0.2 to 0.2.  A major reason that this probability distribution is so flat is due to the large value of the exponent $\sigma = 5.815 \approx 6$.  This is in contrast to the mean field models which have $\sigma = 4$.  For a smaller, probably more realistic volume from the perspective of nuclear collisions, the fluctuations would be even greater.  The magnitude of these fluctuations suggests that it is difficult to probe the properties of the matter very close to the chiral critical point.       

\begin{figure}
\begin{center}\includegraphics[width=4.3in,angle=90]{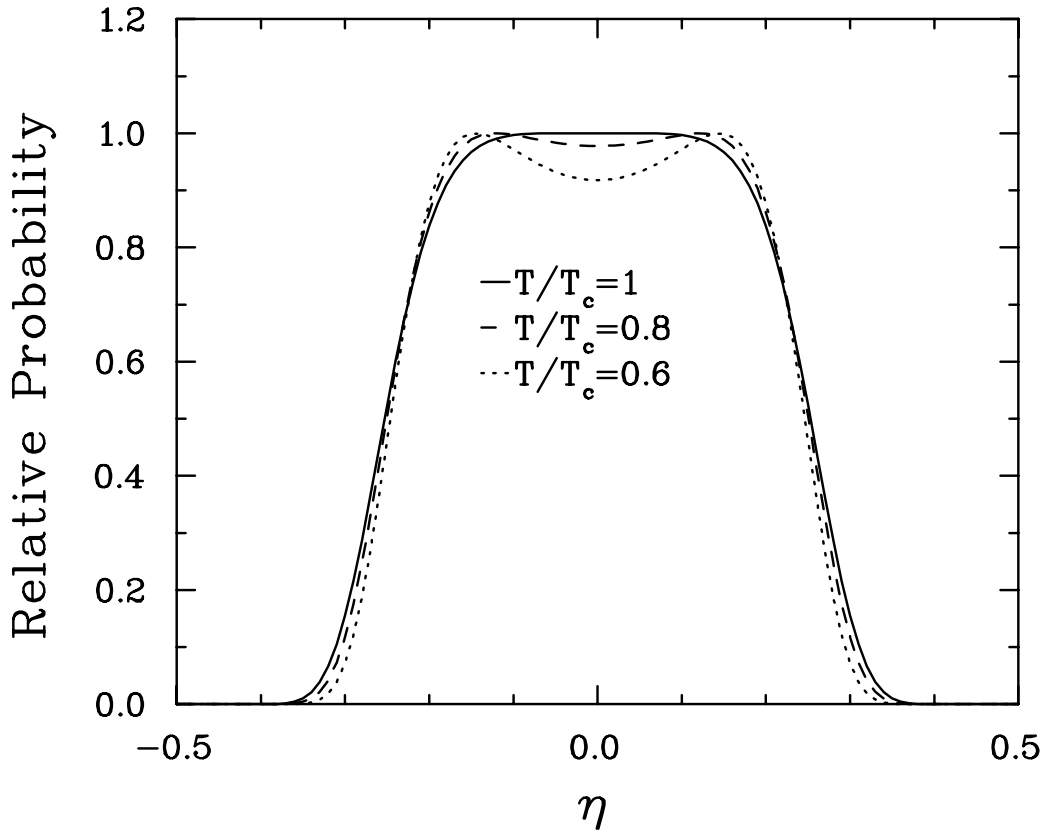}
\caption{The probability to find the system at a particular density relative to the equilibrium densities at phase coexistence.  Three different temperatures are shown.}
\end{center}
\label{probs}
\end{figure}

\newpage

\section{Conclusions}

In this paper we have constructed an equation of state in the vicinity of the chiral critical point.  It incorporates the correct values of the critical exponents and amplitudes.  Since only certain properties of the equation of state are universal, there is some freedom to vary the noncritical functional dependence on temperature and density and to change the parameters in those functions.  The parameterization proposed here matches on to the equation of state at zero baryon density as calculated in lattice gauge theory, and at zero temperature using reasonable extrapolations of dense nuclear matter.  Certainly, refinements and modifications are possible within the present framework.

The Landau theory of fluctuations away from equilibrium states was employed to determine the potential magnitude of the fluctuations one might expect in heavy ion collisions.  The magnitude of these fluctuations is quite large, partly due to finite volume effects but primarily because the critical exponent $\delta$ is much larger than in standard mean field theories.  This flattens the Landau free energy as a function of density away from the equilibrium densities and hence decreases the cost to fluctuate away from them.

In the future it would be highly desirable to have a parameterization of the equation of state that includes not only the behavior near the critical point but also extends to much higher temperatures and densities.  Ultimately, to compare with experimental data, it will be necessary to incorporate this knowledge into dynamical simulations of heavy ion collisions.

\section*{Acknowledgements}

I am grateful to L. Csernai for discussions, and to both him and K. Rajagopal for comments on the manuscript.  This work was supported by the US Department of Energy (DOE) under Grant No. DE-FG02-87ER40328.

\newpage

\section*{Appendix}

Here we review the parameterization of the equation of state near the chiral critical point as constructed in this manuscript for ease of application.  See the text for detailed explanations.

The critical point lies somewhere along the curve
\be
\left(\frac{T}{T_0}\right)^2 + \left(\frac{\mu}{\mu_0}\right)^2 = 1
\ee  
Here $T_0$ and $\mu_0$ are constants.  The pressure at the critical point is estimated from the expression
\be
P = \frac{\pi^2}{90} \left( 16 + \frac{21N_f}{2} \right) T^4 +
\frac{N_f}{18} \mu^2 T^2 + \frac{N_f}{324\pi^2} \mu^4 - CT^2 - B
\ee
where $N_f$ is the number of massless quark flavors.  We use $N_f = 2.5$ to simulate the larger strange quark mass.  The constants $B$ and $C$ are adjusted to represent the results of lattice QCD calculations in the vicinity of the crossover region at $T > 0$ but $\mu = 0$ and to make the pressure a constant along the critical curve.  Then
\be
\frac{\mu_0^2}{T_0^2} = 9\pi^2 \left[ 1 -
\sqrt{ \frac{8}{15} - \frac{32}{45 N_f} } \right]
\approx (6.67173)^2
\ee
\be
C = \frac{N_f \mu_0^2}{18} \sqrt{ \frac{8}{15} - \frac{32}{45 N_f} }
\approx 3.084T_0^2
\ee
\be
B = 0.8 T_0^4
\ee
In particular $P_c \approx 0.749 T_0^4$.  The values of the entropy density, baryon density, and energy density at the critical point are obtained from the above expression for the pressure via thermodynamic identities.  When numerical values are required we use $T_0 = 180$ MeV and thus $\mu_0 = 1209$ MeV.    

The Helmholtz free energy is
\be
f = f_0(t) + f_1(t)\eta + f_2(t)\eta^2 + f_{\sigma}(t)|\eta|^{\sigma} 
\ee
where $\eta = (n-n_c)/n_c$ and $t=(T-T_c)/T_c$.  The value of $\sigma$ is 5.815.  The coefficient functions are
\ba
f_0(t) = \left\{ \begin{array}{ll}
\bar{f}_0(t) -a_- (-t)^{2-\alpha} & \mbox{if $t<0$} \\
\bar{f}_0(t) -a _+ t^{2-\alpha} & \mbox{if $t>0$ }
\end{array} \right.
\ea
\be
f_1(t) = n_c \mu_0 \sqrt{1 - \frac{T_c^2}{T_0^2}(1+t)^2 }
\ee
\ba
f_2(t) = \left\{ \begin{array}{ll}
\bar{f}_2(t)-b_- (-t)^{\gamma} & \mbox{if $t<0$}\\
\; \;\; \bar{f}_2(t) + b_+ t^{\gamma} & \mbox{if $t>0$}
\end{array} \right.
\ea
\be
f_{\sigma} = {\rm constant}
\ee 
The exponents are $\alpha = 0.11$ and $\gamma = 1.24$.  The $\bar{f}_0(t)$ and $\bar{f}_2(t)$ are smooth functions of $t$.  The critical amplitudes are related by
\be
b_+ = \frac{(\sigma - 2) b_-}{5} 
\ee
and
\be
2 a_+ = a_- + \frac{\gamma (\gamma - 1)}{(2-\alpha)(1-\alpha)} 
\left( \frac{2 \, b_-}{\sigma f_{\sigma}} \right)^{\frac{2}{\sigma - 2}} b_- 
\, .
\ee 

From thermodynamic relations the smooth function
\be
\bar{f}_0(t) = \epsilon_c - T_c s_c (1+t)
\ee
to first order in $t$.  The simplest parameterization of the other smooth function is
\be
\bar{f}_2(t)=\thalf b_- \gamma t^2
\ee
The parameters $a_-$ and $b_-$ are
\be
a_- = T_c s_c/(2-\alpha)
\ee
\be
b_- = \frac{\sigma f_{\sigma}}{2-\gamma} 
\left( \frac{\Delta n}{2n_c} \right)^{\sigma - 2}
\ee 
where $\Delta n$ is the discontinuity in the baryon density at $T=0$.  For definiteness we use $\Delta n = n_c/3$ and $f_{\sigma} = 5 P_c \approx 3.745 T_0^4$.


\newpage

\begin{thebibliography}{99}

\bibitem{asakawa89} M. Asakawa and K. Yazaki, Nucl. Phys. {\bf A504}, 668 (1989).

\bibitem{berges98} J. Berges and K. Rajagopal, Nucl. Phys. {\bf B538}, 215 (1999).

\bibitem{scavenius01}
O. Scavenius, A. M\`ocsy, I. N. Mishustin, and D. H. Rischke, Phys. Rev. C 
{\bf 64}, 045202 (2001).

\bibitem{barducci} A. Barducci, R. Casalbuoni, S. De Curtis, R. Gatto, and G. Pettini, Phys. Lett. {\bf B231}, 463 (1989); Phys. Rev. D {\bf 41}, 1610 (1990); A. Barducci, R. Casalbuoni, G. Pettini, and R. Gatto, {\it ibid.} 
{\bf 49}, 426 (1994).

\bibitem{halasz98} M. A. Halasz, A. D. Jackson, R. E. Shrock, M. A. Stephanov, and J. J. M. Verbaarschot, Phys. Rev. D {\bf 58}, 096007 (1998).

\bibitem{hatta02} Y. Hatta and T. Ikeda, Phys. Rev. D {\bf 67}, 014028 (2003).

\bibitem{antoniou02} N. G. Antoniou and A. S. Kapoyannis, Phys. Lett. 
{\bf B563}, 165 (2003).

\bibitem{fodor02} Z. Fodor and S. D. Katz, J. High Energy Phys. 03 (2002) 014; {\it ibid.} 04 (2004) 050.

\bibitem{ejiri03} S. Ejiri, C. R. Allton, S. J. Hands, O. Kaczmarek, F. Karsch, E. Laermann, and C. Schmidt, Prog. Theor. Phys. Suppl. {\bf 153}, 118 (2004).

\bibitem{forcrand03}
Ph. de Forcrand and O. Philipsen, Nucl. Phys. {\bf B642}, 290 (2002);
{\bf B673}, 170 (2003);  Nucl. Phys. Proc. Suppl. {\bf 129}, 521 (2004); J. High Energy Phys. 11 (2008) 012.

\bibitem{gavai05} R. V. Gavai and S. Gupta, Phys. Rev. D {\bf 71}, 114014 (2005).

\bibitem{stephanov}
M. Stephanov, Prog. Theor. Phys. Suppl. {\bf 153}, 139 (2004);
Int. J. Mod. Phys. A {\bf 20}, 4387 (2005); PoS(LAT2006)024.

\bibitem{MohantyQM}
B. Mohanty, preprint, arXiv:0907.4476.

\bibitem{Shuryak}
M. A. Stephanov, K. Rajagopal, and E. Shuryak, Phys. Rev. Lett. {\bf 81}, 4816 (1998); Phys. Rev. D {\bf 60}, 114028 (1999).

\bibitem{Hatta}
Y. Hatta and M. A. Stephanov, Phys. Rev. Lett. {\bf 91}, 102003 (2003); {\bf 91}, 129901(E) (2003).

\bibitem{small}
B. Berdnikov and K. Rajagopal, Phys. Rev. D {\bf 61}, 105017 (2000). 
 
\bibitem{nongaussian}
M. A. Stephanov, Phys. Rev. Lett. {\bf 102}, 032301 (2009).

\bibitem{Randrup}
J. Randrup, Phys. Rev. C {\bf 79}, 054911 (2009).

\bibitem{Goodman}
A. L. Goodman, J. I. Kapusta, and A. Z. Mekjian, Phys. Rev. C {\bf 30}, 851 (1984).

\bibitem{attract}
C. Nonaka and M. Asakawa, Phys. Rev. C {\bf 71}, 044904 (2005).

\bibitem{fluidexpts}
A. C. Flewelling, R. J. Defonseka, N. Khaleeli, J. Partee and D. T. Jacobs, J. Chem. Phys. {\bf 104}, 8048 (1996); C. A. Ramos, A. R. King and V. Jaccarino, Phys. Rev. B {\bf 40}, 7124 (1989).

\bibitem{Zinnbook} J. Zinn-Justin, {\it Quantum Field Theory and Critical Phenomena}, Clarendon Press, Oxford, 3rd edition, 1996.

\bibitem{Guida} R. Guida and J. Zinn-Justin, Nucl. Phys. {\bf B489} [FS], 626 (1997).

\bibitem{Grant} C. Grant and J. Kapusta, Phys. Rev. C {\bf 32}, 663 (1985).

\bibitem{nuc1} P. M\"oller, W. D. Myers, W. J. Swiatecki, and J. Treiner, Atomic Data Nuclear Data Tables {\bf 39}, 225 (1988); W. D. Myers and W. J. Swiatecki, Phys. Rev. C {\bf 57}, 3020 (1998).

\bibitem{nuc2} J. P. Blaizot, Phys. Rep. {\bf 64}, 171 (1980); J. P. Blaizot, J. F. Berger, J. Decharge, and M. Girod, Nucl. Phys. {\bf A591}, 435 (1991); Dao T. Khoa, G. R. Satchler, and W. von Oertzen, Phys. Rev. C {\bf 56}, 954 (1997); D. H. Youngblood, H. L. Clark, and Y. W. Lui, Phys. Rev. Lett. {\bf 82}, 691 (1999).

\bibitem{KapGale} J. I. Kapusta and C. Gale, {\it Finite Temperature Field Theory}, Cambridge University Press, Cambridge, 2nd edition, 2006.

\bibitem{HIreviews} J. W. Harris and B. M\"uller, Ann. Rev. Nucl. Part. Sci. {\bf 46}, 71 (1996); S. Das Gupta and G. D. Westfall, Physics Today {\bf 46}, 34 (1993).

\bibitem{Hagedorn} R. Hagedorn, {\it Cargese Lectures in Physics}, Vol. 6, ed. E. Schatzmann, Gordon and Breach, 1973.

\bibitem{excluded} J. I. Kapusta and K. A. Olive, Nucl. Phys. {\bf A408}, 478 (1983).

\bibitem{statmodels} J. Cleymans and K. Redlich, Phys. Rev. Lett. {\bf 81}, 5284 (1998); Phys. Rev. C {\bf 60}, 054908 (1999); J. Cleymans, H. Oeschler, K. Redlich, and S. Wheaton, {\it ibid.} {\bf 73}, 034905 (2006).

\bibitem{Fodor} Y. Aoki, Sz. Borsanyi, S. Durr, Z. Fodor, S. D. Katz, S. Krieg, and K. K. Szabo, JHEP {\bf 0906}, 088 (2009). 

\bibitem{HotQCD} A. Bazavov, {\it et al.}, Phys. Rev. D {\bf 80}, 014504 (2009). 

\bibitem{Boyd} G. Boyd, J. Engels, F. Karsch, E. Laermann, C. Legeland, M. Lutgemeier, and B. Peterson, Nucl. Phys. {\bf B469}, 419 (1996).

\bibitem{Tsquare} R. D. Pisarski, Prog. Theor. Phys. Suppl. {\bf 168}, 276 (2007).

\bibitem{musquare} J. Ellis, J. I. Kapusta, and K. A. Olive, Nucl. Phys. 
{\bf B348}, 345 (1991).

\bibitem{Landau} E. M. Lifshitz and L. P. Pitaevskii, {\it Statistical Physics}, 3rd edition (Pergamon, New York, 1980), Part 1.

\end{thebibliography}
\end{document}